\documentclass[useAMS,usenatbib]{mn2e}
\usepackage{psfig}

\usepackage{times}

\def\lsim{ \lower .75ex\hbox{$\sim$} \llap{\raise .27ex \hbox{$<$}} }
\def\gsim{ \lower .75ex \hbox{$\sim$} \llap{\raise .27ex \hbox{$>$}} }

\title[Fast TeV variability in blazars] 
{TeV variability in blazars: how fast can it be?}

\author[Ghisellini et al.]
{G. Ghisellini$^1$\thanks{E--mail: gabriele.ghisellini@brera.inaf.it}, 
 F. Tavecchio$^1$,  G. Bodo$^2$ and A. Celotti$^3$ \\
$^1$INAF -- Osservatorio Astronomico di Brera, via E. Bianchi 46, I--23807
Merate, Italy\\
$^2$INAF -- Osservatorio Astronomico di Torino, strada dell'Osservatorio, 20,  I--10025
Pino Torinese, Italy \\
$^3$S.I.S.S.A., V. Beirut 2--4, I-34014 Trieste, Italy
}

\begin{document}



\maketitle

\begin{abstract} Recent Cerenkov observations of the two BL Lac
  objects PKS 2155--304 and Mkn 501 revealed TeV flux variability by a
  factor $\sim$ 2 in just 3--5 minutes.  Even accounting for the
  effects of relativistic beaming, such short timescales are
  challenging simple and conventional emitting models, and call for
  alternative ideas.  We explore the possibility that extremely fast
  variable emission might be produced by particles streaming at
  ultra--relativistic speeds along magnetic field lines and inverse
  Compton scattering any radiation field already present. This
  would produce extremely collimated beams of TeV photons.  While the
  probability for the line of sight to be within such a narrow cone of
  emission would be negligibly small, one would expect that the
  process is not confined to a single site, but can take place in many very
  localised regions, along almost straight magnetic lines.  
A possible astrophysical setting realising these conditions 
is magneto--centrifugal acceleration of beams of particles.
In this scenario, the variability timescale would not be
  related to the physical dimension of the emitting volume, but might
  be determined by either the typical duration of the process
  responsible for the production of these high energy particle beams
  or by the coherence length of the magnetic field.  It is predicted
  that even faster TeV variability -- with no X--ray counterpart --
  should be observed by the foreseen more sensitive Cerenkov
  telescopes.
\end{abstract}


\section{Introduction}

In the standard framework, the overall non-thermal energy distribution
of blazars is produced within a relativistic jet closely aligned to
the line of sight. While the origin of the flux variability is not
known, for a variability timescales $t_{\rm var}$ the general
causality argument imposes the limit $R \lsim ct_{\rm var} \delta$ on the
typical dimension $R$ of a quasi--spherical emitting region, where
$\delta $ is the Doppler factor of the radiating plasma. This
constrain can be bypassed if the dimension of the region along the
line of sight is much smaller than the other two dimensions.  Such a
configuration is the natural one arising 
in shocks forming within the flow (internal
shock model, Spada et al. 2001, Guetta et al. 2004).  In this scenario
the minimum variability timescale is related to the Schwarzschild
radius $R_{\rm s}$ of the central black hole, $t_{\rm var}>R_{\rm s}/c$, 
due to the cancellation of the bulk Lorentz factors\footnote{Two 
shells separated by $R_0\approx R_{\rm s}$,
having a width of the same order, and having bulk Lorentz factors
differing by a factor 2 will catch up at a distance $z\sim \Gamma^2 R_{\rm s}$,
radiating for an observed time $t_{\rm var}
\sim z/(c \Gamma^2)\sim R_{\rm s}/c$.}.

Unprecedented ultra--fast variability at TeV energies has been
recently detected from the blazars Mkn 501 (Albert et al. 2007) and
PKS 2155--304 (Aharonian et al. 2007), on timescales as short as
$t_{\rm var}\sim$ 3--5 minutes.  And in the latter source (at
$z=0.116$) the variable high energy radiation corresponded to an
observed (isotropic) luminosity $L \sim 10^{47}$ erg s$^{-1}$, which
was dominating the broadband emission.

As pointed out by Begelman, Fabian \& Rees (2008) -- the observations
of ultra--fast variability strongly challenges the above framework in
both geometrical configurations.  In the case of a quasi--spherical
region, the short variability timescale implies that the source is too
compact (see Begelman et al. 2008) unless extreme values for $\delta$
($> 100$) are assumed, at odds with -- among other beaming indicators
-- the relatively low velocity estimated for the knots in their pc--scale
jets (Piner et al. 2008; but see Ghisellini et al. 2005 for a possible
solution).  In the internal shock scenario, $t_{\rm var}$ is totally
inconsistent with the typical black hole masses hosted in blazar
nuclei.

In principle there is no lower limit on the dimension of an
emitting region
and thus variability timescales could be decoupled from the typical
minimum scale of the system: the average emission, typically varying
on timescales $t_{\rm var}\sim 10^4$ s, could still be produced over
volumes comparable to the jet size, while sporadic, ultra--fast flares 
could originate
in very localised regions.  However, as the flux of the ultra--fast
flares was comparable to the bolometric one, a further condition would
have to occur (such as an extremely efficient radiation mechanism, a
high Lorentz factor for the emitting plasma, a particular geometry).
It is thus meaningful to wonder whether there is any robust 
{\it physical limit} to the observed duration and luminosity of flares.

In this work we tackle this question in the context of leptonic
emission models, i.e. the observed high energy radiation is produced,
via inverse Compton, by relativistic leptons. 
We first consider a completely ideal case which maximises the
effects of relativistic beaming showing that, under particular
conditions involving beams of highly relativistic emitting particles,
no observationally interesting limit holds.  Then the astrophysical
feasibility of such an ideal case is examined, and we propose a more
specific setting which seems an ideal environment to produce such
narrow beams.

\section{An idealised limit to fast TeV variability}

Relativistic amplification of the emitted radiation is the key
physical process on which the standard model for blazars is based.
Typically it is postulated that high--energy electrons ($\gamma >10^5$) 
move with random directions within the emitting region which,
in turn, is propagating with a bulk Lorentz factor of $\Gamma \sim 10$
at a small angle with respect to the line of sight.

However, if the highly relativistic electrons were almost co--aligned
in a narrow {\it beam} (as considered by, e.g., Aharonian et
al. 2002, Krawczynski 2008) we can achieve a more efficient situation -- 
in terms of detected emission -- for observers aligned with the beam.
Before assessing the
physical feasibility of such a configuration let us consider the
consequences on the observed emission.

The energy loss for (standard) inverse Compton (IC) scattering of an
electron with Lorentz factor $\gamma = (1-\beta^2)^{-1/2}$ embedded in
a radiation field of energy density $U_{\rm r}$ is (e.g. Rybicki \&
Lightman 1979):
\begin{equation}
P \, = \,\dot\gamma m_{\rm e} c^2\, =\, 
{4\over 3} \sigma_{\rm T} c \gamma^2 U_{\rm r}, 
\label{pe}
\end{equation}
%
where $\sigma_{\rm T}$ is the Thomson cross section and the seed
photon field is assumed isotropically distributed.
In Eq. \ref{pe} $P$ represents the power emitted by the electrons,
while the power received by an observer depends on the viewing angle:
within the cone $1/\gamma$ time is Doppler contracted by the factor
$(1-\beta)$ and the isotropic equivalent power $P_{\rm iso}$ is
enhanced by the factor $(1-\beta)^{-1}$, as the radiation is
collimated in a solid angle $\Delta \Omega =2\pi (1-\beta)$.  The two
effects combine to yield a maximum observed power
\begin{equation}
P_{\rm iso, max} \, = (1-\beta)^{-2} P\, = (1+\beta)^2 \gamma^4 P
\sim {16 \over 3} \sigma_{\rm T} c \gamma^6 U_{\rm r}.
\label{pr}
\end{equation}
If the electron and the observer remain "aligned" for a time longer
than the radiative cooling time $t_{\rm c}$, radiation will be seen
for
\begin{equation}
t_{\rm r} \, = (1-\beta) t_{\rm c} \, = \, 
{ 3 m_{\rm e} c \over 
4 \sigma_{\rm T}(1+\beta) \gamma^3 U_{\rm r} }.
\label{tr}
\end{equation}

Let us estimate how many electrons $N$ are required in order to
observe $P_{\rm iso, max}\sim 10^{47}L_{47}$ erg s$^{-1}$ in the TeV
range\footnote{We adopt the notation $Q=10^x Q_x$, with cgs units.}.
  For simplicity we first consider the case where the source has no
  bulk motion and the seed photons are isotropically distributed.
This requires $\gamma\ge 10^6$, in order to produce TeV  photons.
  Then
\begin{equation}
N\, =\, {10^{47} L_{47} \over P_{\rm iso, max}} \, =\, 
9.4 \times 10^{59}\, {L_{47} \over \gamma^6 U_{\rm r}} =\, 
9.4 \times 10^{23}\, {L_{47} \over \gamma^6_6 U_{\rm r}},  
\label{mg}
\end{equation}
corresponding to a mass $N m_{\rm e} \sim 0.85 L_{47}/(\gamma^6_6
U_{\rm r})$ milligrams and an energy $E=\gamma N m_{\rm e} c^2=
7.7\times 10^{23} L_{47}/(\gamma^5_6 U_{\rm r})$ erg.

Emission would be observed for a mere $t_{\rm r} \, = 1.5\times
10^{-11}/(\gamma_6^3 U_{\rm r})$ s, during which an electron would
travel for a cooling distance $\Delta R_{\rm c} =\beta c t_{\rm c}=
9\times 10^{11}$ cm towards the observer.

This idealised limit to the shortest time variability observable in
the TeV range, even thought unrealistic, shows that high amplitude,
apparent luminosity variability is physically possible even over
sub--nanosecond timescales.

In the sketched scenario the variability timescales -- unrelated to the
size of the emitting region -- may reflect how long a beam maintains
its coherence and alignment with the observer's line of sight, and/or
the duration of the process responsible to produce such a beam.

\subsection{Energy requirements}

The first issue to be discussed in relation to the idealised case
concerns the feasibility of attaining realistic configurations which
allow this ``streaming scenario" without requiring a large amount of
energy.

If electrons stream along magnetic field lines, 
the latter should maintain
their direction (within a factor $1/\gamma$) for a minimum scale
length, of the order of the electron cooling one ($\sim \Delta R_c
\sim 10^{11}-10^{12}$ cm).  This probably imposes the most severe
constrain, but it is hard to meaningfully quantify it.
If electrons move along non--parallel (or partly curved) magnetic field
lines, at any given time the probability of observing radiation (from
electrons pointing along the line of sight) increases.  On the other
hand, this of course requires more electrons (to account for those not
emitting towards the observer at a given instant).

For a flare during which the observed luminosity $L$ doubles in a time
$\Delta t$, the numbers and total energy of electrons required to
radiate the observed average $L$ are a factor $\Delta t / t_{\rm r}$
larger than what just derived above. As an illustrative example,
the TeV flare of PKS 2155--304 lasted for $\Delta t = 100 \Delta t_2$ s, yielding
$\Delta t / t_{\rm r} = 6.7\times 10^{12} \Delta t_2 \gamma_6^3 U_{\rm
  r}$.  It follows that the total energy which has to be invoked to
sustain the observed average $L=10^{47} L_{47}$ erg s$^{-1}$ for
$\Delta t$ is
\begin{equation}
E\, =\, \gamma N m_{\rm e} c^2 {\Delta t \over t_{\rm r}} \,=\,
5.2\times 10^{36}\, { L_{47}\Delta t_2 \over \gamma^2_6} \quad {\rm erg}.
\end{equation}
Note that $E$ is independent of $U_{\rm r}$ (as expected, since only
electron energies are involved). The term $\gamma^{-2}$ enters
through the solid angle of the beamed radiation.

The inferred energetics would be not particularly demanding, but it
corresponds to a single particle beam. As the probability that such a
beam is oriented along the line of sight with an accuracy of order
$1/\gamma$ is exceedingly small, the existence of many beams along
field lines, whose directions cover a sizeable fraction $f$ of the jet
opening angle $\theta_{\rm j}$, is mandatory for the process to be of
any astrophysical relevance.

If $A$ is the number beams -- each subtending a solid angle
$\Delta\Omega_{\rm b} = 2\pi [1-\cos(1/\gamma)]$ -- with directions
within the jet solid angle $\Delta\Omega_{\rm j} = 2\pi
(1-\cos\theta_{\rm j})$,
\begin{equation}
f\, \approx\, A\,  {\Delta \Omega_{\rm b} \over \Delta\Omega_{\rm j}}
\, \sim \,  {A \over (\gamma \theta_{\rm j})^2} \, \sim \, 10^{-10} \,{A \over 
(\gamma_6 \theta_{\rm j, -1})^2}.
\end{equation}
$f$ can be in principle roughly estimated by the ``duty cycle" of the
high energy ultra--fast flares, namely the fraction of the
observational time during which flares are visible.  
Therefore, if one
ultra--fast flare is detected during an observing time interval
$T_{\rm obs}$,
\begin{equation}
  f\, =\, {\Delta t \over T_{\rm obs}} \, =\, 
  10^{-3} \, {\Delta t_2 \over T_{\rm obs, 5}} \, \to \,
  A\, \sim 10^7 \, (\gamma_6 \theta_{\rm j, -1})^2 \, {\Delta t_2 \over T_{\rm obs, 5}}
\end{equation}
where $T_{\rm obs}\sim 10^5$ s has been assumed (likely a lower limit).  

Consequently the total energetics, $E_{\rm tot}$, which also accounts
for beams not pointing at the observer amounts to   
\begin{equation}
E_{\rm tot} \, = \, A E\, =\, 
5.2\times 10^{43}\,  {\Delta t_2  \over T_{\rm obs, 5}} \,
 L_{47} \Delta t_2 (\theta_{\rm j, -1})^2   
\quad {\rm erg}
\label{etot}
\end{equation}
independently of $\gamma$. This corresponds to the bulk energy of
leptons responsible of flares. Modelling of the average spectral energy
distribution of TeV blazars, and in particular of PKS 2155--304 (see
e.g. Foschini et al. 2007; Celotti \& Ghisellini 2007; Ghisellini \&
Tavecchio 2008) imply jet powers of the order $L_{\rm jet}\sim
10^{45}$ erg s$^{-1}$, largely exceeding $E_{\rm tot}/\Delta t$. We
conclude that from the energetics point of view, the proposed
mechanism is not demanding.

\subsection{Partial isotropization of pitch angles and synchrotron emission}

The mechanism responsible for the acceleration of particles along
field lines could in principle favour a distribution of pitch angles
$\psi$ peaked at small values, of the order $\sim 1/\gamma$ (see
Section 3).

It is also likely though that small disturbances in the field
configuration result in values $\psi$ greater than $1/\gamma$.  The
solid angle of emission is then $\Delta \Omega_{\rm b} =2\pi
(\psi/\gamma) \gg \pi/\gamma^2$: the IC 
power emitted by the beam will
spread over a larger solid angle (implying a reduced observed flux),
but this effect will be compensated by photons emitted along the line
of sight from other beams. 


When the electrons acquire a non--vanishing pitch angle $\psi$,
they also emit synchrotron radiation. 
It is interesting to evaluate the corresponding flux.
%
Like the IC radiation, the
synchrotron power $P_{\rm s, iso}$, received by an ``aligned" observer,  
is amplified by a factor $(1-\beta\cos\psi)^{-1}$ with respect to the
emitted one, $P_{\rm s, e}$. 
The radiation is collimated within a solid
angle of the order $\Delta \Omega_{\rm s}\sim 2\pi \sin\psi/\gamma$,
\begin{eqnarray}
P_{\rm s, iso} \, &=& \, P_{\rm s, e}\, \, {4\pi \gamma \over 2\pi \sin\psi} \, \,
{ 1\over 1-\beta\cos\psi} \nonumber \\
 &=&\,
4\sigma_{\rm T} c U_B \gamma^3\beta^2  {\sin\psi \over 1-\beta\cos\psi} 
\label{psiso}
\end{eqnarray}
This is maximised at $\sin\psi=1/\gamma$ at a value
%
\begin{equation}
P_{\rm s, iso, max}\, =\, 4\sigma_{\rm T} c U_B \gamma^4\beta^2,
\quad (\sin\psi =1/\gamma), 
\label{psmax}
\end{equation}
which is a factor $2\gamma^2$ larger than for an electron with pitch
angle $\psi=90^\circ$.  
With respect to Eq. \ref{pr}, the ratio of
observed and emitted power is smaller by a factor $\gamma^2$, as
synchrotron emission has an extra pitch--angle dependence ($\propto
1/\gamma^2$ for $\psi\sim 1/\gamma$).

To summarise. Only IC radiation (and no synchrotron) is observed from
leptons perfectly aligned with the magnetic field lines, and in turn
with the observer. For pitch angles of the order $1/\gamma$, the
synchrotron flux received from a single electron is maximised, yet the
ratio of the received powers (IC/synchrotron) is a factor $\sim
\gamma^2$ larger (cfr Eq. \ref{pr} with Eq. \ref{psmax}) than the
corresponding ratio in the isotropic pitch angle case. In other words,
radiation from streaming electrons will produce effects more
pronounced in the IC than in the synchrotron branch of the spectral
distribution.


\begin{figure}
\psfig{figure=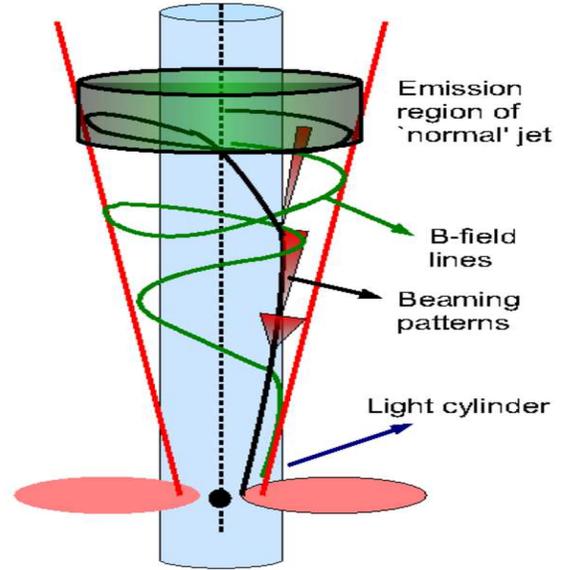,width=9cm,height=9cm}
\caption{Schematic representation of the proposed scenario.  At some
  distance from the black hole and accretion disc part of the jet
  power is dissipated in a region moving with a bulk Lorentz factor
  10--30, producing radiation through ``standard'' synchrotron and
  self--Compton emission.  Magneto--centrifugally accelerated electrons
  are bound to stream along magnetic field lines, initially oriented
  with different directions.  They would reach the light cylinder at
  different locations where the electrons reach their maximum energy.
  Beyond the light cylinder field lines wound--up since they cannot
  move rigidly any longer. Occasionally, magnetic field lines close to
  the light cylinder would point towards the observer.
  In these cases one can then
  detect the relativistically boosted radiation resulting from the IC
  scattering by the magneto--centrifugally accelerated electrons
  off the seed photons produced by the ``normal'' jet and/or 
  by the accretion flow.}
\label{jet}
\end{figure}

\section{A possible astrophysical setting: magneto--centrifugal
  acceleration}


A natural astrophysical mechanism producing a configuration similar to
the one proposed, where highly relativistic electrons stream along
magnetic field lines, is centrifugal acceleration
(e.g. Rieger \& Mannheim 2000, Osmanov et al. 2007, Rieger \&
Aharonian 2008).

Several models invoking centrifugal acceleration assume
that magnetic field lines rigidly rotate at a fraction of the black
hole rotational velocity.  A charged test particle, injected at the
base and co--rotating with the field (as ``bead--on--wire") will
experience the centrifugal force and will be accelerated with an
efficiency that increases as the particle approaches the light
cylinder (Machabeli \& Rogava 1994; Machabeli et al. 1996; Gangadhara
\& Lesch 1997; Rieger \& Mannheim 2000; Osmanov et al. 2007).  The
energy boost will be limited by radiative losses due to IC 
process and/or by the breakdown of the ``bead--on--wire"
approximation, when the Coriolis force -- tearing the particle off the
field line -- exceeds the Lorentz force.  The effective limiting
mechanism depends on the accretion disc luminosity: the former
(latter) will dominate for higher (lower) radiation energy densities.
Maximum electron Lorentz factors around $\gamma =10^8$ can be attained
under reasonable conditions when, as in TeV emitting BL Lacs, the
accretion disc is radiatively very inefficient (Osmanov et al. 2007).
If, besides the accretion disc, there are other sources of cooling
photons, the maximum $\gamma$ will be smaller. 
In TeV BL Lacs the radiation produced by the ``normal" jet region can in 
some cases dominate (but not by a huge factor) the cooling. 
We have checked that in such conditions the maximum $\gamma$--factor
reaches values as large as $10^6$, as required to produce TeV radiation.
Protons suffer less from radiative
losses and their maximum energy is limited by the breakdown of 
the ``bead--on--wire" approximation.
As a result, protons and electrons can achieve
comparable energies in such a case.

Before and during the acceleration three important effects will concur
to decrease the final electron pitch angle:
\begin{enumerate}
\item the magneto--centrifugal force increases only the parallel
  component of the particle momentum;
\item the magnetic field decreases with distance: therefore the pitch
  angle of the accelerated electron will also decrease, as in a
  magnetic bottle;
\item at the beginning of the acceleration, the electron is likely
  sub--relativistic due to strong radiation losses since close to the
  accretion disc the radiation and magnetic fields are most
  intense (see below).
\end{enumerate}
In the following we examine these effects in some details:
\vskip 0.2 true cm
\noindent
{\bf Magneto--centrifugal acceleration --}
The force vector can be decomposed into two components, parallel and
perpendicular to the magnetic field line.  During a gyro--orbit, the
perpendicular one acts half of the time in favour and half against the
electron motion, with a null average effect.  Only the parallel
momentum of the particle will then be increased in the process.
\vskip 0.2 true cm
\noindent {\bf Adiabatic invariant --} Since the electron moves along
divergent magnetic field lines, its pitch angle will decrease. A
simple estimate can be made using the adiabatic invariant in the form
\begin{equation}
{(p'_\perp)^2 \over B'} \, = \, {\rm constant} \, \to p'_\perp \, 
\propto \, (B')^{1/2},
\label{pperp}
\end{equation}
where $p'_\perp\equiv \gamma'_\perp\beta'_\perp$ is the dimensionless
transverse electron momentum in the gyro--frame.  A Lorentz
transformation in the lab frame yields:
\begin{equation}
\gamma \, = \, \gamma_\parallel \gamma^\prime_\perp, \qquad
\beta_\perp \, = \, {\beta'_\perp \over \gamma_\parallel },
\label{gtot}
\end{equation}
from which 
\begin{equation}
\tan\psi \, = \, {\beta_\perp \over\beta_\parallel} \, \to \, 
\sin\psi \, = {p'_\perp \over \gamma\beta}. 
\label{psi}
\end{equation}
As $p'_\perp $ will decrease owed to the decrease of the $B$--field
(Eq. \ref{pperp}) while $\gamma$ will increase thanks to
acceleration, the pitch angle will decrease. 
Thus even though
initially an electron is mildly relativistic in the gyro--frame
(i.e. $p'_\perp \gsim 1$) this ensures that the final pitch angle is
of the order of $1/\gamma$.  
\vskip 0.2 true cm
\noindent
{\bf Radiative cooling --}
For an electron with initial large pitch angle and large $p'_\perp$ the
decreasing magnetic field might not be sufficient to yield a final
small $\psi$. However, close to the accretion disk, the magnetic field
is large, implying severe radiation losses. While synchrotron losses
do not affect the pitch angle (synchrotron photons radiated by a
relativistic electron are emitted along the electron velocity
direction), they limit $p'_\perp$ (e.g. $p'_\perp$ reaches
trans--relativistic values for $B\sim 100$ G at a scale of $R\sim
10^{15}$ cm).  A similar effect is produced by IC scattering, as in
the electron rest frame virtually all seed photons are seen as coming
from the forward direction and the scattering cross section is
azimuthally symmetric (both in the Thomson and in the Klein Nishina
regime).

\subsection{The spectrum: qualitative considerations}

The equilibrium distribution of particles, solution of the kinetic
equation including the acceleration and cooling terms, can be
described by a power--law with typical slope $N(\gamma)\propto
\gamma^{-n}$, with $n=3/2$ (Rieger \& Aharonian 2008). For an isotropic pitch
angle distribution the corresponding IC spectrum is of the form
$\propto\nu^{-(n-1)/2}\propto \nu^{-1/4}$ (if the seed photons are 
monochromatic and their energy density is constant along the electron beam).
If the seed photons are distributed as 
$F(\nu_{\rm s})\propto \nu_{\rm s}^{-\alpha_s}$, with $\alpha_s>1/4$,
the slope of the scattered spectrum is the same of the slope of the seeds,
i.e. $\alpha_s>1/4$, and somewhat steeper at high frequencies if Klein--Nishina
effects are important.
This is the limit when there is a large ensemble 
of beams covering a wide range of directions.

%


The other limit is when only the radiation from a  single beam can be
observed. This comprises particles with a range of $\gamma$ and,
correspondingly, of pitch angles (cfr Eq. \ref{psi}): lower energy
electrons, predominant at the start of the acceleration process, will
have the largest pitch angles; the most energetic electrons, with the
smallest pitch angles, will tend to be located at the end of the
accelerating zone and will mostly contribute to the emission in the
TeV band.

The observed spectral slope of the radiation from a single beam can be
qualitatively estimated for a beam along a (straight) field line
perfectly aligned with the line of sight, and a particle energy
power--law distribution $N(\gamma)\propto \gamma^{-n}$.

Let us discuss first the case of monochromatic seed photons,
of frequency $\nu_0$.
The up--scattered photons will have an average frequency $\nu\propto
\gamma^2\nu_0$ for scatterings in the Thomson regime, 
and $\nu\propto \gamma$ in the Klein--Nishina one.

If the energy density of the seeds are roughly the same for electrons
of low and high $\gamma$, the resulting spectral slope will correspond to
\begin{eqnarray}
L(\nu) \propto  N(\gamma)P_{\rm iso, max} {d\gamma \over d\nu}  &\propto& 
\nu^{(5-n)/2} \, \quad{\rm Thomson} \nonumber \\
&\propto& \nu^{(6-n)} \qquad{\rm Klein~Nishina},
\label{slope}
\end{eqnarray}
as $P_{\rm iso, max}\propto \gamma^6$ for a single
electron (Eq. \ref{pr}).  
As above, if the seed photons are distributed as 
$F(\nu_{\rm s})\propto \nu_{\rm s}^{-\alpha_s}$
the slope of the scattered spectrum is the same of the slope of the seeds,
and somewhat steeper at high frequencies if Klein--Nishina
effects are important.

The observed spectral slope will depend on the
distributions of the emitting beams.
In any case, at TeV energies, we expect a slope very similar to the
slope of the seed photon distribution, steepening at the highest 
frequencies because of Klein--Nishina effects.

As for the seed photons, we will have always at least two
components: the radiation coming from the (inefficient) accretion
flow (see Mahadevan 1997 for illustrative examples of disc spectra) 
and the radiation produced by the ``normal" active region of the jet. 
Their relative importance depends on their luminosities and spectra, the
distance of the electron beams, and the bulk velocity of the active
region of the jet. For typical values (i.e. disc luminosities $\sim 10^{41}$
erg s$^{-1}$, jet comoving luminosities $\sim 10^{43}$ erg s$^{-1}$,
and beams located at $\sim 10^{17}$ cm from a disc of size $\sim 10^{16}$ cm)
the jet radiation dominates.
In this case the spectrum produced by the electron beam is expected to
be very similar to the SSC spectrum of the jet.

\section{Discussion and observational tests}

We have suggested that ultra--fast TeV variability could originate
from particles ``streaming" along magnetic field lines, namely beams
of highly relativistic electrons with very small pitch angles that
occasionally point towards the observer, giving rise to flare events.

These leptons could inverse Compton scatter synchrotron photons
produced by the population of electrons responsible for the broad band
radiation detected most of the observing time, characterised by
variability timescales of a few hours. The latter, ``normal" jet 
emission also comprises a synchrotron self--Compton TeV component, 
expected to vary
coherently with the synchrotron one at X--rays frequencies.

Usually, the streaming particles would point in directions off the
line of sight, but changes in the magnetic lines orientation result in
a non zero probability that they become closely aligned with it. 
The probability for this to occur depends on the geometry,
the degree of coherence of the magnetic field and the total number of
particle beams (pointing in any direction). The latter number can be
estimated from the (admittedly still poorly determined) duty cycle of
the ultra--fast variability events: the required total energy is not
demanding.

Within this scenario there is in principle no astrophysical
interesting limit on how fast TeV variability can be. Variations the
order of $10^{47}$ erg s$^{-1}$ in the apparent luminosity can occur even over
sub--nanosecond timescales.  However a typical minimum variability
timescale could be estimated for a specific geometrical setting.  For
illustration, consider a configuration where magnetic field lines,
before reaching the light cylinder, rigidly rotate at some velocity
$\beta_{\rm B} c$ and the emission region is located at some distance
$z$ from the black hole.  Particles travelling along a given field
line with pitch angle $\sim 1/\gamma$, will emit in a particular
direction for a time:
\begin{equation}
t_{\rm var} \, = \, {2\over \gamma}\, {2\pi \theta_{\rm j} z \over 
\beta_{\rm B} c } 
\,\sim \, 0.5\, {\theta_{\rm j,-1} z_{\rm 16} \over 
\beta_{\rm B} \gamma_6} \, \, {\rm s},
\label{sec}
\end{equation}
where $2\pi \theta_{\rm j}z /\beta_{\rm B} c$ is the rotational period
of the field line and $2/\gamma$ is the fraction of the rotational period
during which the beaming cone subtends the line of sight. 
Thus in this geometrical situation the minimum variability 
timescale is expected to be of the order of a second. 
This simply refers to a single field line, aligned (within a
factor $1/\gamma$) for at least one cooling length (i.e. for $\sim
10^{12}$ cm) -- if this is not the case $t_{\rm var}$ would be
shorter. It also assumes a single bunch of electrons whose emission
can be observed: $t_{\rm var}$ would be longer if we can detect the 
radiation from particles streaming on other adjacent field lines.

The variability timescale of these ultra--fast events is not related
to the typical dimension of the emitting region and depends on the
duration of the acceleration phase and on the time interval over which 
the magnetic field lines are aligned with the line of sight.

Magneto--centrifugal acceleration scenario can easily produce beams of
electrons with pitch angles of the order $1/\gamma$.  This can be
achieved if initially (i.e. at the base of the jet) the particles are
not relativistic, as indeed radiation losses ensure. 

The general scenario where ``needle beams'' of very energetic
electrons with small pitch angles can account for ultra--fast TeV
variability bear some relevant consequences that can be
observationally tested:

\begin{itemize}

\item Although in principle TeV variability timescales could be
  extremely short, in the proposed astrophysical setting a typical
  minimum value can be of the order of a second (Eq. \ref{sec}).  As a
  consequence, more and more sensitive Cerenkov telescopes and
  arrays should detect faster and faster flux variability. 
  Peak fluxes need not to be smaller for shorter events.

\item No correlation between X--ray and TeV flux is expected during
  ultra--fast flares, as synchrotron emission from the streaming
  particles is weaker than the inverse Compton one by a factor
  $\gamma^2$.  Ultra--fast TeV variability should resemble the
  phenomenology of ``orphan flares'' (as detected from the TeV BL Lac
  1959+650; Krawczynski et al. 2004).  On the contrary, the
  synchrotron and inverse Compton fluxes produced by the normal jet 
  should vary in a correlated way.
  Furthermore the normal jet synchrotron flux is likely to always dominate
  over the streaming particle synchrotron component and thus no
  ultra--fast events should be observed in the X--ray band.

\item Variability should be faster at higher inverse Compton
  frequencies, as they are produced by the higher energy electrons
  which also have the smallest pitch angles. This also implies that
  the observed flux has to be produced by a smaller number of streaming
  particle bunches, namely higher energy flares should be rarer.

\item TeV spectra, during ultra--fast variability, are expected to be
similar to less active phases, as observed in the flaring state of
PKS 2155--304 (Aharonian et al. 2007).

\end{itemize}

\section*{Acknowledgments}
This work was partly financially supported by a 2007 COFIN-MIUR grant.

\end{document}